\documentclass[11.5pt]{irsreport}
\usepackage{epsfig}

\title{IRS-TR 11001:  Temporal Responsivity Variations on the Red Peak-Up 
Sub-Array}

\author{
G.C. Sloan (1) ~\& D.A. Ludovici (2) \thanks{ (1) Infrared Spectrograph 
Science Center, Cornell University, (2) Department of Physics, West 
Virginia University; NSF REU Research Assistant, Astronomy Department, 
Cornell University} }

\date{3 August, 2011}

\runninghead{IRS-TR 11001 - Red PU Variations}

\begin{document}

\maketitle

\begin{abstract}

Over the course of the cryogenic mission of the {\it Spitzer 
Space Telescope}, the responsivity of the Red Peak-Up 
sub-array on the Infrared Spectograph (IRS)  varied by 
$\sim$2\%, based on an analysis of five standard stars.   The 
sensitivity dropped 1.7\% after the first 14 IRS campaigns, 
then climbed back up 1.0\% later in the mission.  The 
uncertainty in these measurements is better than $\sim$0.3\%.  
The random variations in the Peak-Up photometry of the 
standard stars has a gaussian distribution of width 
$\sim$2\%, similar to the magnitude of the systematic 
temporal variations.

\end{abstract}

\section{Introduction} % Sec. 1

The cryogenic mission of the {\it Spitzer Space Telescope}
(Werner et al.\ 2004) lasted from August 2003 to May 2009.
In that time, the Infrared Spectrograph (IRS; Houck et al.\
2004) made repeated observations of several standard stars.
Those fainter than $\sim$ 1 Jy at 12~\mum\ could be observed 
both spectroscopically and on the Peak-Up (PU) subarrays on 
the Short-Low (SL) module.  It was highly desirable to do so 
for two reasons.  First, using the target for self peak-up 
reduced the likelihood of possible pointing errors, because 
as long as the coordinates were good to within several
arcseconds, the on-board PU algorithm would find the source 
and shift it with its full precision to the center of the 
spectrographic slits.  With an offset star, any errors with 
the coordinates of {\it either} star translated into a 
pointing error.  Second, the PU observations gave us an 
independent photometric measurement of the target, allowing 
us to monitor the stability of both the standard stars and
the instrument.  If a standard varied in isolation compared 
to the others, then it probably is a variable star.  If all 
of the standards varied in concert, variation in system 
response is the likely culprit.

\section{Observations} % Sec. 2

Table~1 lists the five stars observed most frequently with 
the Red Peak-Up sub-array during the cryogenic {\it Spitzer} 
mission.  These stars were faint enough to not saturate the 
Red PU array, which would have resulted in poorer pointings 
for spectroscopy, and they were brighter than $\sim$ 150 mJy 
at 16~\mum.  For stars below this limit, we used the Blue PU 
sub-array, which gave slightly more accurate pointings due to 
the smaller point-spread function (PSF).  Figure 1 
illustrates the responsivity of the two PU sub-arrays, which 
peak at approximately 16~\mum\ (Blue) and 22~\mum\ (Red), 
along with the actual spectra of the standards considered here.

\bigskip
\noindent \begin{tabular}{lllrrr} % Table 1
\multicolumn{6}{c}{\bf Table 1---Peak-Up Acquisition Observations} \\
\hline
{\bf Standard} & {\bf Spec.} & {\bf Observed in} & {\bf Self-PU} & 
  \multicolumn{2}{c}{\bf Red PU images} \\
{\bf star} & {\bf class} & {\bf campaigns} & {\bf AORs} & 
  {\bf rejected} & {\bf used} \\
\hline
HR~6348      & K1 III & P--61 (intermittently) &   84 &  5 & 163 \\
HD~166780    & K4 III & P--61 (intermittently) &   41 &  3 &  79 \\
HD~173511    & K5 III & 5--61                  &  132 &  9 & 217 \\
$\alpha$ Lac & A1 V   & 1--58 (intermittently) &   28 &  2 &  54 \\
$\delta$ UMi & A1 V   & P--19 (intermittently) &   43 &  3 &  83 \\
\hline
\end{tabular}
\bigskip

\begin{figure} % Fig. 1 % [ht!]
  \begin{center}
     \epsfig{file=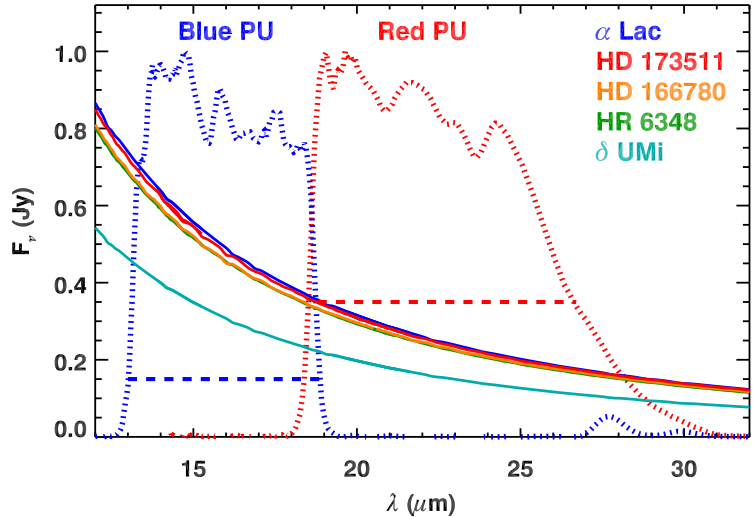, width=10cm}
  \end{center}
\caption{
%\small
---The normalized responsivities of the Blue and Red PU 
sub-arrays, for a bias voltage of 2.0 volts, as operated 
during the {\it Spitzer} mission (dashed curves).  The 
horizontal dashed lines show the photometric levels at which 
the PU sub-arrays begin to saturate.  Spectra of the five
standards considered here are also plotted.}
\end{figure}

The PU data were obtained in two sets of three consecutive 
images, which were read using double-correlated sampling and 
then combined and analyzed on the spacecraft.  After the 
first set, or ``acquisition'' set, the centroid was 
determined and the telescope moved to shift the target to a 
well-calibrated portion of the PU array known as the 
``sweet-spot''.  The three ``sweet-spot'' images were then
combined and analyzed as before.  The second centroid was 
then used to shift the target to the first requested 
spectroscopic slit.  Thus, each PU observation results in two 
combined images and two independent photometric measurements.

Early in the mission, we were very conservative about 
saturating any pixels in the PU sub-arrays, which would 
reduce the accuracy the centroiding algorithm and introduce
possible pointing errors.  Consequently, the observations of 
HD~173511 used offset PUs through IRS Campaign 4, and the 
corresponding astronomical observing requests (AORs, 20 in
all) have no PU photometry for this source.
  
\section{Analysis} % Sec. 3

The analysis begins with S18.18 pipeline output from the
{\it Spitzer} Science Center (SSC).  We use the ``acqr.fits''
files, which are more processed than the raw ``bcd.fits''
files available in previous pipeline versions.  The ``bcd''
images can suffer from what is known as ``jailbarring''.  The 
images are read one row of 128 pixels at a time, with 32 
reads of four-pixel groups by four parallel analog-to-digital 
converters (ADCs).  Sometimes the dark current from the four 
ADCs drifts, resulting in vertical stripes in the images.  
The ``acqr'' images delivered by the SSC correct this 
artifact.

We combined the three images from one position using median
filtering, and then measured the signal from the target using
simple aperture photometry in two ways.  We use a narrow
aperture of radius four pixels and a corresponding sky
radius of eight pixels, and a broad aperture of radius
seven pixels with a sky radius of 14 pixels.  In both cases,
the sky is measured in the outer annulus and removed from
the photometry inside the inner circle.  The two apertures
produce similar results, but because the photometry in the
narrow apertures shows less noise, we concentrate on those
results here.

In rare cases, the photometry from the ``acquisition'' and 
``sweet spot'' images disagree with each other, most likely
because of latent images in the SL detector from previous
observations.  In those cases, the acqusition image is far
more likely to produce spurious results.  We have filtered
the data by using only the second (sweet-spot) image when
the difference between the two images is greater than
1.5\% of their mean.  As a result, only a handful of data
remain outside an envelope of $\sim\pm$5\%.

\begin{figure} % Fig. 2
  \begin{center}
     \epsfig{file=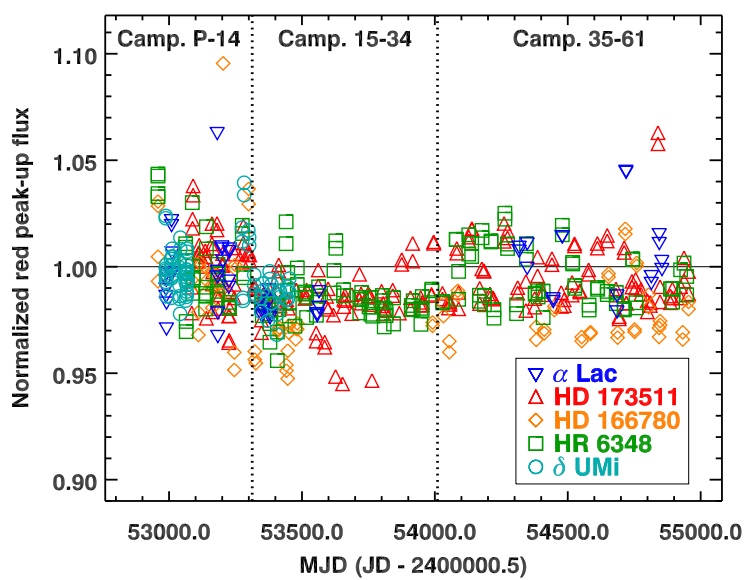, width=10cm}
  \end{center}
\caption{
%\small
---The photometry of the five standards most frequently observed 
in the Red PU sub-array, normalized to their mean response in 
the period from Campaigns P through 14.  The responsivity drops
$\sim$2\% during the middle portion of the cryogenic mission and
then climbs $\sim$1\% climb in later campaigns.  
The vertical dashed lines separate the individual campaigns 
into the three sets we use to correct the photometry for
these systemic shifts. (MJD = modified Julian date = JD $-$ 
2400000.5.)}
\end{figure}

\bigskip
\begin{tabular}{lrrlll} % Table 2
\multicolumn{6}{c}{\bf Table 2---Normalized Mean Photometry} \\
\hline
{\bf Target}  & {\bf Campaigns} & {\bf N} & {\bf mean} & 
{\bf $\sigma$} & {\bf u} \\
\hline
HR~6348       &  P--14 &  37 & 1.0    & 0.0187 & 0.0031 \\
              & 15--34 &  64 & 0.9840 & 0.0106 & 0.0013 \\
              & 35--61 &  62 & 0.9849 & 0.0134 & 0.0017 \\
\\
HD~166780     &  P--14 &  24 & 1.0    & 0.0291 & 0.0059 \\
              & 15--34 &  19 & 0.9649 & 0.0085 & 0.0019 \\
              & 35--61 &  36 & 0.9784 & 0.0142 & 0.0024 \\
\\
HD~173511     &  P--14 &  52 & 1.0    & 0.0158 & 0.0022 \\
              & 15--34 &  78 & 0.9845 & 0.0119 & 0.0013 \\
              & 35--61 &  87 & 0.9784 & 0.0153 & 0.0016 \\
\\
$\alpha$ Lac  &  P--14 &  24 & 1.0    & 0.0190 & 0.0039 \\
              & 15--34 &  12 & 0.9820 & 0.0046 & 0.0013 \\
              & 35--61 &  18 & 1.0060 & 0.0183 & 0.0043 \\
\\
$\delta$ UMi  &  P--14 &  56 & 1.0    & 0.0124 & 0.0017 \\
              & 15--34 &  25 & 0.9872 & 0.0069 & 0.0014 \\
              & 35--61 &   0 & ...... & ...... & ...... \\
\\
\hline
\\
Combined mean &  P--14 & 193 & 1.0    & 0.0181 & 0.0013 \\
              & 15--34 & 198 & 0.9826 & 0.0103 & 0.0007 \\
              & 35--61 & 203 & 0.9929 & 0.0149 & 0.0010 \\
\hline
\end{tabular}
\bigskip

Figure 2 illustrates the resulting photometry, normalized for
each source to its mean over the period from Campaign P in
the Science Verification (SV) phase to Campaign 14 during 
normal operations.  All five stars follow the same trend of
a decline in signal of $\sim$2\% from early in the mission
to about half-way through, followed by a slight $\sim$1\% 
increase in signal from then to the end of the cryogenic
mission.  Because all five stars vary in unison, we conclude
that we are dealing with variation in the system.  We will
assume that the issue is a subtle change in responsivity.
Any additive offset would have been removed by the sky
subtraction.

The spread in the data at any given time is on the order of a 
few percent, about as large as the apparent trends.
Consequently, we do not attempt any kind of polynomial fit to 
the responsivity as a function of time.  Instead, we have 
separated the cryogenic mission into three phases, each 
containing approximately equal numbers of photometric
measurements and corresponding roughly to the overall
breaks in the apparent trends.

Table 2 presents the means for each standard for the three 
phases of the mission, the standard deviation ($\sigma$) and 
the uncertainty in the mean ($u = \sigma/\sqrt{N}$).  The 
last set of data in the table gives the combined means of all 
five standards during the mission, weighted to account for 
the number of observations.  Our overall photometric 
uncertainty is $\sim$0.1\%, compared to shifts in system 
responsivity of 1.7\% down in the middle campaigns and 1.0\% 
back up for later campaigns.

\section{Discussion} % Sec. 4

\begin{figure} % Fig. 3 
  \begin{center}
     \epsfig{file=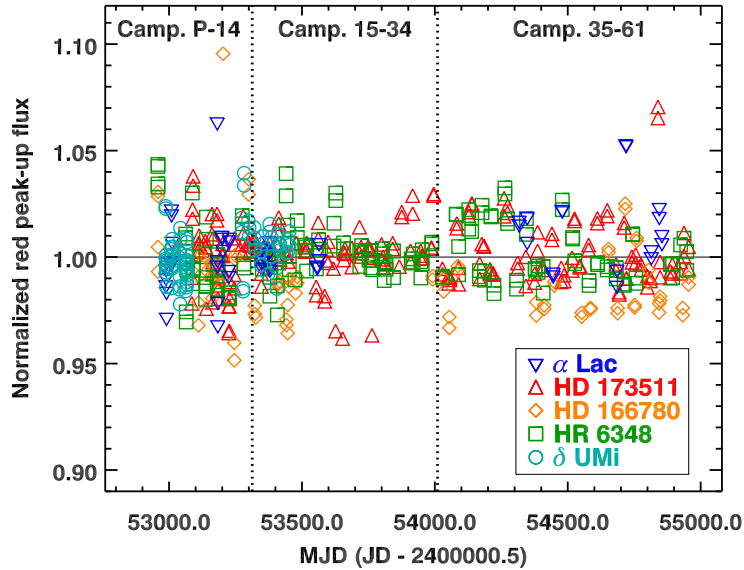, width=10cm}
  \end{center}
\caption{
%\small
---The Red PU photometry of the five standards from Fig.\ 2, 
after applying the recommended corrections upward of 1.70\% 
for Camp.\ 15--34 and 0.70\% for Camp.\ 35--61.}
\end{figure}

Based on this analysis, we recommend divisive corrections to
Red PU photometry of 0.9826 ($\pm$0.0007) for IRS Campaigns 
15--34 and 0.9929 ($\pm$0.0010) for IRS Campaigns 35--61.  
These corrections bring all data into line with
the early IRS campaigns.  For most users of the IRS, this 
level of precision for red PU photometry is probably 
unnecessary, but we will apply these corrections in future 
technical reports and papers describing the spectrophotometric 
calibration of the IRS and its application to the sample of 
spectroscopic standards observed during the cryogenic mission.

Ludovici et al.\ (2011) presented an earlier analysis using
S18.7 data at the AAS meeting in Seattle.  They found
corrections within $\sim$0.5\% of the values presented here
with different filtering for outliers and slightly different 
breaks between campaigns (the middle group was Camp.\ 
16--34).  Analysis of the S18.7 data more consistent with the 
steps described in Section 3 produce results within 
$\sim$0.15--0.3\% of our recommended corrections above, 
depending on the detailed assumptions.  These differences 
give a reasonable estimate in our uncertainty.

\bigskip
\begin{tabular}{lc} % Table 3
\multicolumn{2}{c}{\bf Table 3---Detrended means} \\
\hline
{\bf Target}  &  {\bf Detrended mean}  \\
\hline
HR~6348       &  1.0013 $\pm$ 0.0011 \\
HD~166780     &  0.9890 $\pm$ 0.0023 \\
HD~173511     &  1.0015 $\pm$ 0.0010 \\
$\alpha$ Lac  &  1.0043 $\pm$ 0.0024 \\
$\delta$ UMi  &  1.0014 $\pm$ 0.0012 \\
\hline
\end{tabular}
\bigskip

Figure 3 illustrates the impact of the recommended 
corrections on the data.  The overall trends apparent in 
Figure 2 have largely been removed.  Table 3 presents the 
detrended means for the five standards.  The means were
determined after the data were first normalized to the 
signal for each star in Campaigns P--14, then detrended with 
the recommended divisive corrections given above (0.9826 and 
0.9929).  For three standards (HR~6348, HD~173511, and 
$\delta$~UMi), the difference between the corrected means 
and unity are close to the uncertainty in the mean.  

For HD~166780, however, the mean is 0.989, which is a 
more significant difference.  Comparing the data up to and
after Campaign 14, we have 1.0 $\pm$ 0.0059 vs.\ 0.9843 $\pm$
0.0017, which is a difference of 0.0157 $\pm$ 0.0061, 
indicating a 1.6\% drop in emission over the {\it Spitzer}
mission with a 2.6-$\sigma$ level of confidence.  We cannot 
claim conclusively that HD~166780 is a variable star, but 
this change in flux does raise some concerns.

The photometric behavior of $\alpha$ Lac is also troublesome,
but while a similar analysis suggests a 0.7\% increase in 
brightness, the confidence level is only 1.3~$\sigma$, making
it impossible to draw any conclusions.

\begin{figure}[hb!] % Fig. 4
  \begin{center}
     \epsfig{file=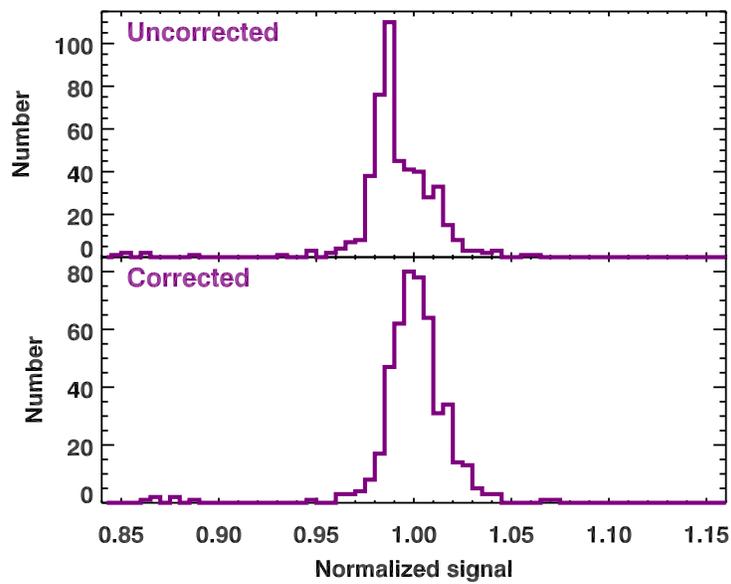, width=10cm}
  \end{center}
\caption{
%\small
---The combined Red PU photometry for HR~6348, HD~173511, and 
$\delta$~UMi after normalizing the data for each source to 
the mean of Campaigns P--14.  The top panel shows the PU
signals before we have corrected the data from later 
campaigns for the temporal responsivity variations.  The
bottom panel shows the effect of the correction.  The data
plotted in this figure include outliers filtered from
previous analysis.}
\end{figure}

Figure 4 illustrates how the corrections improve the
distribution of the Red PU photometry from the three most
stable sources:  HR~6348, HD~173511, and $\delta$~UMi.
The histograms include {\it all} of the data from these
three sources, including the images filtered in the 
analysis above (by rejecting the acquisition image when it
disagrees with the sweet-spot image by more than 1.5\%).
The lower distribution is clearly more gaussian in behavior,
even if the standard deviation has improved only slightly, 
from 2.2\% to 2.0\%.  This random distribution is similar in
spread as the more systematic temporal variations measured
above.

\end{document}